\documentclass[aps, prb, twocolumn, showpacs, amsmath, amssymb, 
superscriptaddress, floatfix, longbibliography]{revtex4-2}

\usepackage[dvipsnames]{xcolor}

\definecolor{p-channel}{HTML}{445def}
\definecolor{c-channel}{HTML}{fa2217}
\definecolor{d-channel}{HTML}{580772}

\DeclareUnicodeCharacter{2032}{$\prime$}

\usepackage{mathtools}
\usepackage{braket}
\usepackage{textcomp,gensymb}
\usepackage{dsfont}
\usepackage{comment}

\usepackage{graphicx}
\usepackage[export]{adjustbox}

\usepackage{hyperref}
\usepackage[capitalize]{cleveref}
\usepackage{csquotes}

\usepackage{algpseudocode}
\usepackage{bbold}
\usepackage[mode=buildnew]{standalone}
\usepackage{makerobust}

\usepackage{pifont}
\usepackage{tikz}

\makeatletter
\def\convertto#1#2{\strip@pt\dimexpr #2*65536/\number\dimexpr 1#1}
\makeatother
\newcommand{\measurewidths}{{%
  \color{red}
  :::
  TEXT \convertto{in}{\the\textwidth} in
  :::
  COLUMN \convertto{in}{\the\columnwidth} in
  :::
  HEIGHT(ex) \convertto{pt}{1ex} pt
  :::
  WIDTH(em) \convertto{pt}{1em} pt
  :::
}}

\newcommand*\circled[1]{\tikz[baseline=(char.base)]{
    \node[shape=circle,draw,inner sep=1pt] (char) {#1};}}
\MakeRobustCommand\circled

\newcommand{\bvec}[1]{\boldsymbol #1}

\newcommand{\dd}{\mathrm{d}}

\newcommand{\makeauthor}[2]{\newcommand{#1}[1]{{%
  \sffamily\color{#2}{%
    \bfseries\begingroup\escapechar=-1\edef\x{\endgroup\string#1}\x:%
  } ##1}}%
  \MakeRobustCommand#1}
\makeauthor{\lk}{BurntOrange}
\makeauthor{\jb}{ForestGreen}
\makeauthor{\jh}{NavyBlue}
\makeauthor{\ts}{magenta}
\makeauthor{\sr}{blue}

\begin{document}

\title{Rashba spin-orbit coupling in the square lattice Hubbard model:\\
a truncated-unity functional renormalization group study}

\author{Jacob Beyer}
\affiliation{School of Physics, University of Melbourne, Parkville,
    VIC 3010, Australia}
\affiliation{Institute for Theoretical Solid State Physics,
    RWTH Aachen University, 52062 Aachen, Germany}
\affiliation{JARA Fundamentals of Future Information Technology,
    52062 Aachen, Germany}
\affiliation{Institute for Theoretical Physics, University of Würzburg,
    Am Hubland, 97074 Würzburg, Germany}
\author{Jonas B.~Profe}
\affiliation{JARA Fundamentals of Future Information Technology,
    52062 Aachen, Germany}
\affiliation{Institute for Theory of Statistical Physics,
    RWTH Aachen University, 52062 Aachen, Germany}
\author{Lennart Klebl}
\affiliation{JARA Fundamentals of Future Information Technology,
    52062 Aachen, Germany}
\affiliation{Institute for Theory of Statistical Physics,
    RWTH Aachen University, 52062 Aachen, Germany}
\author{Tilman Schwemmer}
\affiliation{Institute for Theoretical Physics, University of Würzburg,
    Am Hubland, 97074 Würzburg, Germany}
\author{Dante M. Kennes}
\affiliation{JARA Fundamentals of Future Information Technology,
    52062 Aachen, Germany}
\affiliation{Institute for Theory of Statistical Physics,
    RWTH Aachen University, 52062 Aachen, Germany}
\affiliation{Max Planck Institute for the Structure and Dynamics of Matter,
    Center for Free Electron Laser Science, 22761 Hamburg, Germany}
\author{Ronny Thomale}
\affiliation{Institute for Theoretical Physics, University of Würzburg,
    Am Hubland, 97074 Würzburg, Germany}
\author{Carsten Honerkamp}
\affiliation{Institute for Theoretical Solid State Physics,
    RWTH Aachen University, 52062 Aachen, Germany}
\affiliation{JARA Fundamentals of Future Information Technology,
    52062 Aachen, Germany}
\author{Stephan Rachel}
\affiliation{School of Physics, University of Melbourne, Parkville,
    VIC 3010, Australia}

\date{\today}

\begin{abstract}
    The {\it Rashba--Hubbard model} on the square lattice is the paradigmatic case for
    studying the effect of spin-orbit coupling, which breaks spin and inversion symmetry, in a correlated electron system.
    We employ a truncated-unity variant of the functional renormalization group which
    allows us to analyze magnetic and superconducting instabilities on equal footing.
    We derive phase diagrams depending on the strengths of Rasbha spin-orbit coupling,
    real second-neighbor hopping and electron filling.
    We find commensurate and incommensurate magnetic phases which compete with $d$-wave 
    superconductivity.
    Due to the breaking of inversion symmetry, singlet and triplet components mix;
    we quantify the mixing of  $d$-wave singlet pairing with $f$-wave triplet pairing.
\end{abstract}

\maketitle

\section{Introduction}

Topological superconductors are amongst the most desirable materials, due to their
huge potential for future information processing technology and fault-tolerant
quantum computation\,\cite{nayak-08rmp1083}.
Such materials can be designed as hetero or hybrid structures via proximity
effect\,\cite{mourik_signatures_2012,nadj-perge_observation_2014,palacio-morales_atomic-scale_2019,kim_toward_2018,schneider_topological_2021},
or the topological superconductivity arises as an intrinsic many-body instability
in a correlated electron system. The latter is usually associated with
odd-parity or spin-triplet pairing, exemplified through the archetypal chiral
$p$-wave state\,\cite{Read-99prb8084,ivanov01prl268}.
Triplet superconductors are rare in nature\,\cite{fesete1,fesete2,wolf-22prl167002},
but it was appreciated in the past years that spin-orbit coupling (SOC) is beneficial
to stabilize triplet superconductivity.
As a consequence, there is growing interest in correlated materials involving
heavy elements or hybrid- and heterostructures in which the inversion symmetry
is broken at the interface.

Prominent material realizations involve interfaces of transition metal oxides such as
LaAlO$_3$/SrTiO$_3$.
Interestingly, magnetic moments seem to be omnipresent at the interface.
Experimental reports include both ferromagnetic and antiferromagnetic
order\,\cite{brinkman-07nm493}, but also spiral magnetism seems to be
possible\,\cite{banerjee-13np626}, hinting at the role of Rashba spin-orbit coupling.
A particularly remarkable result is the observation of the coexistence of magnetism
and superconductivity\,\cite{li-11np762,dikin-11prl056802}.
It was further shown that there are effective ways of tuning the strength of the
Rashba coupling at the interface by an applied electric
field\,\cite{caviglia2010tunable}.
The tunability of Rashba spin-orbit coupling was also reported in the related iridate
heterostructure LaMnO$_3$/SrIrO$_3$ by varying the growth
conditions\,\cite{suraj2020tunable}.

The heavy-fermion superconductor CeCoIn$_5$/YbCoIn$_5$ constitutes another example
where Rashba coupling and electron correlations
coexist\,\cite{mizukami-11np849,shimozawa-14prl156404}.
The material has the intriguing property that the Rashba spin-orbit strength
can be tuned by varying the number of layers in the YbCoIn$_5$ blocks.
Superconductivity is mediated by magnetic fluctuations\,\cite{stock-08prl087001},
underlining the importance of analyzing magnetic and superconducting
instabilities simultaneously.

CePt$_3$Si is one of the best-studied instances of a strongly correlated material
with inversion symmetry breaking that becomes superconducting at low
temperatures\,\cite{bauer-04prl027003,yanase-07jpsj043712,smidman-17rpp036501}.
In CePt$_3$Si, the absence of a mirror plan induces a Rashba spin-orbit coupling.
Experiments hinted at the unconventional nature of the superconductor and found
line nodes in the spectrum, which could be explained through singlet-triplet
mixing due to spin-orbit coupling\,\cite{yanase-08jpsj124711,smidman-17rpp036501}.

One of the most surprising developments in the past few years are the recent results
on the overdoped high-temperature superconductor Bi$_2$Sr$_2$CaCu$_2$O$_{8+\delta}$
(Bi2212)\,\cite{Gotlieb-revealing-2018}.
Cuprates are a prototype system where the degree of strong correlations leads to
a complex interplay of competing interactions;
however, Rashba spin-orbit coupling was assumed to be negligible.
In the recent work, a non-trivial spin texture with spin-momentum locking was observed
in Bi2212, one of the most-studied cuprate superconductors.
These results challenge the standard modeling for cuprates involving Hubbard models,
and emphasize the need for extending correlated electron models with  Rashba coupling.

Motivated by these and other material
examples\,\cite{smidman-17rpp036501,https://doi.org/10.48550/arxiv.2207.02806},
the role of Rashba spin-orbit coupling on the phase diagram of Hubbard models
has attracted considerable interest in the past
decade\,\cite{shigeta_superconducting_2013,laubach-14prb165136,greco_mechanism_2018,ghadimi-19prb115122,rachel_wolf,greco-20prb174420,kawano-22arXiv2208.09902,Wang_2016}.
Several of these theoretical works have used mean-field methods,
random-phase approximation, weak-coupling methods or other approximate approaches.
A major challenge is to account for particle-hole instabilities
(e.g. spin and charge density waves) and particle-particle instabilities
(i.e., superconductivity) and analyze them on equal footing. 

In this paper, we employ the truncated-unity functional renormalization group (TUFRG)
method, and demonstrate that it represents an efficient and powerful method to tackle
the described task.
In \cref{sec:modelmethod} we introduce model and method.
In \cref{sec:results} we present our results consisting of phase diagrams
established through analysis of particle-hole and particle-particle instabilities.
We conclude in \cref{sec:conclusion}.

\section{Model and Method}
\label{sec:modelmethod}

\subsection{Square lattice Hubbard-Rashba Model}

We consider a model of electrons on the two-dimensional square lattice. The non-interacting part of the Hamiltonian is composed of two terms,  spin-independent hopping $H_\mathrm{kin}$ and Rashba SOC $H_\mathrm{SOC}$,
\begin{equation}
\begin{aligned}
H_\mathrm{kin} &{}= \sum_{ij\sigma} t_{ij}c^\dagger_{i\sigma}c^{\vphantom{\dagger}}_{j\sigma} \,,\\
H_\mathrm{SOC} &{}= i\alpha\,\sum_{ij\sigma\sigma'}t_{ij}\big(\hat{\bvec\sigma}\times\bvec r_{ij}\big)_{z}^{\sigma\sigma'}
c^\dagger_{i\sigma}c^{\vphantom{\dagger}}_{j\sigma'} \,.
\end{aligned}
\end{equation}
Here, $c^{(\dagger)}_{i\sigma}$ annihilates (creates) an electron with spin $\sigma$ at lattice site $i$. We allow nonzero hopping among nearest ($t_{\braket{ij}} = t$) and next-nearest ($t_{\braket{\braket{ij}}} = t'$) neighbors. The SOC strength is controlled by $\alpha$, which couples via the cross product of the Pauli matrices $\hat{\bvec\sigma} = (\hat\sigma_x,\hat\sigma_y,\hat\sigma_z)^\mathrm{T}$ with the bond vectors $\bvec r_{ij}$ connecting sites $i$ and $j$.

We complement the non-interacting Hamiltonian $H_0=H_\mathrm{kin} + H_\mathrm{SOC}$ with a purely local Hubbard interaction,
\begin{equation}
H_\mathrm{int} = \frac{U}2\sum_{i\sigma} n_{i\sigma} n_{i\bar\sigma} \,,
\end{equation}
with $n_{i\sigma}=c^\dagger_{i\sigma}c^{\vphantom{\dagger}}_{i\sigma}$ the occupation number and $U$ the interaction strength. Motivated by insights gained from weak coupling renormalization group studies\,\cite{wcRG,rachel_wolf}, we treat the interacting part $H_\mathrm{int}$ with the functional renormalization group (FRG). A short introduction is given below.

\subsection{Functional Renormalization Group}
The unbiased nature of the FRG allows us to explore the realm of possible phase transitions of the Rashba-Hubbard model.
Using FRG, we can obtain information on particle-particle (superconducting) and particle-hole (magnetic, charge, etc.)
instabilities on equal footing.
Furthermore, the general scope of FRG trivially allows for the mixing of singlet and triplet components of the
superconducting order parameter, which are expected to occur in this model due to
the unconventional $d$-wave order of the Hubbard model in conjunction with the Rashba-SOC.
While there have been no previous studies of the repulsive square lattice
Rashba--Hubbard model using FRG,
the limiting case $\alpha = 0$ --- the square lattice Hubbard model --- has been most thoroughly
covered\,\cite{Halboth_2000,PhysRevB.89.035126,husesalm,lichtenstein2017,honerkamp_magnetic_2001,vilardi_antiferromagnetic_2019,tagliavini_multiloop_2019} and is used as a benchmark in \cref{sec:validation}.
Previous FRG studies including Rashba-SOC have been carried out on the triangular lattice for (i) a model with attractive $U$\,\cite{schober2016} and (ii) a materials-oriented model of twisted bilayer PtSe\textsubscript{2}\,\cite{klebl_moire}.

\begin{figure}
    \centering
    \includegraphics{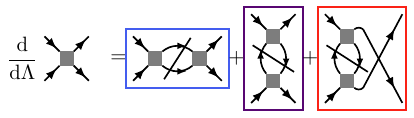}
    \caption{Diagrammatic representation of the FRG flow equation. The two-particle nodes represent the effective interaction
    at scale $\Gamma^{(4)}_\Lambda$ while the connecting lines are a propagator $G^\Lambda$ and its
    derivative $\frac{\dd}{\dd\Lambda}G^\Lambda$ denoted by the line intersecting the loops.
    We indicate the three different two-particle irreducible diagram classes (channels):
    particle-particle \textcolor{p-channel}{$P$}, crossed particle-hole \textcolor{c-channel}{$C$},
    and direct particle-hole \textcolor{d-channel}{$D$}.
    Each is associated with one distinct transfer momentum.
    } 
    \label{fig:diags}
\end{figure}

Derivations of the FRG equations can be found in Refs.\,\onlinecite{metzner-salmhofer-honerkamp-2012,dupuis_nonperturbative_2021,platt-hanke-thomale-2013,behak} among others, here we only give the briefest overview of our chosen approximations: The diagrammatic representation of our single-loop, non-$SU(2)$ flow-equation is given in \cref{fig:diags}. During the calculation we neglect higher loop orders, the flow equations for the self-energy $\frac{\dd}{\dd\Lambda}\Gamma^{(2)}$ as well as the three-particle interactions $\frac{\dd}{\dd\Lambda}\Gamma^{(6)}$, focusing on the effective two-particle interaction $\Gamma^{(4)}$. As regulator for the two-point Greens function we choose the $\Omega$-cutoff\,\cite{husesalm}, where $G^{\Lambda}(\omega, \bvec k) = \frac{\omega^2}{\omega^2+\Lambda^2}G(\omega, \bvec k)$. Moreover, we restrict ourselves to zero temperature $T = 0$. As we disregard frequency dependencies and the self-energy feedback\,\cite{PhysRevB.67.174504} of the effective interaction we are able to evaluate all occurring Matsubara frequency integrations analytically.

To reduce computational complexity we employ the truncated unity extension to the FRG\,\cite{lichtenstein2017,PhysRevB.79.195125,PhysRevB.85.035414}. Importantly, this retains momentum conservation at the vertices, broken in $N$-patch schemes\,\cite{PhysRevB.63.035109}, but required to accurately capture spin-momentum locking. For a more complete technical discussion of the approximations and numerical implementations we refer the reader to Ref.\,\onlinecite{behak}.

To obtain a prediction for the low-energy two-particle interaction $\Gamma^{(4)}_{\mathrm{eff}}$ without the artificial scale we solve the differential equation of \cref{fig:diags} starting at infinite (large compared to band-width) scale $\Lambda_{\infty}$ and integrate successively towards zero. When encountering a phase-transition, associated elements of $\Gamma^{(4)}$ will diverge driven by the diverging susceptibilities and the truncation at the four-point vertex $\Gamma^{(4)}$ introduced above is no longer valid. We therefore terminate the integration at this critical scale $\Lambda_\mathrm c$ and analyze the effective two particle interaction $\Gamma^{(4)}_{\Lambda_\mathrm c}$ to determine the type of ordering associated with the transition.

\subsection{Analysis of Results}
As the TUFRG scheme naturally splits the vertices into particle-particle ($P$) and particle-hole ($C$, $D$) channels (see \cref{fig:diags}), we can start the analysis by finding the channel that dominantly contributes to the divergence of $\Gamma^{(4)}$. Thereafter, we calculate interacting susceptibilities ($\chi$) in the subspace of transfer momenta $\bvec q$ of the leading vertex elements in the respective channel (see \cref{eqn:susc_pp,eqn:susc_ph}). We subsequently perform an eigen-decomposition of the susceptibilities to determine weights (eigenvalues) and order parameters (eigenvectors) of the respective transitions. In the case of superconducting instabilities at $\bvec q=0$, we additionally solve a linearized gap equation and obtain a Fermi surface projected order parameter as well as one in the full Brillouin zone (see e.g.\ Refs.~\onlinecite{behak,klebl_moire,klebl_competition}). The order parameters serve as a starting point for further analysis, e.g.\ competition of singlet- and triplet superconductivity, or charge-density-wave (CDW) vs. spin-density-wave (SDW) instabilities.

\subsubsection{Particle-particle instabilities}
We calculate the particle-particle susceptibility from the effective interaction at the final scale $\Gamma^{(4)}_{\Lambda_\mathrm c}$ projected to the $P$-channel:
\begin{equation}
    \includegraphics[valign=c]{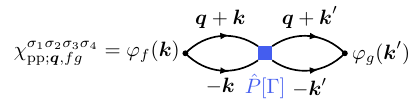}.
    \label{eqn:susc_pp}
\end{equation}
The functions $\varphi_{f,g}(\bvec k)$ are the basis functions used in the truncated unity expansion (\enquote{formfactors}). As the channel projected vertex is given in formfactor space, we insert unities of the form $\delta(\bvec k - \bvec k') = \sum_f \int \! \dd\bvec k \varphi^{*}_f(\bvec k) \varphi^{\phantom{*}}_{f} (\bvec k')$ into \cref{eqn:susc_pp} to carry out the calculation. The particle-particle loops in \cref{eqn:susc_pp} and the following \cref{eqn:lingap} are evaluated at the critical scale $\Lambda_\mathrm c$. In case the leading transfer momentum of $\chi_\mathrm{pp}$ is at $\bvec q=0$ (as expected for instabilities that are not pair-density waves\,\cite{PhysRev.108.1175}), we additionally solve the following linearized gap equation for the superconducting order parameter $\Delta_{\sigma\sigma'}(\bvec k)$:
\begin{equation}
    \includegraphics{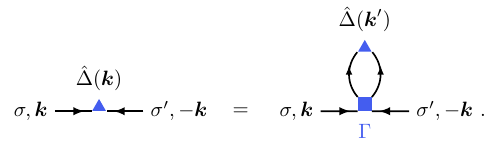}
    \label{eqn:lingap}
\end{equation}
As noted in Refs.\,\onlinecite{klebl_moire,klebl_competition,behak}, the eigenproblem presented in \cref{eqn:lingap} is non-Hermitian. So we resort to a singular value decomposition and obtain left and right singular vectors\,\footnote{%
    Resorting to a singular value decomposition instead of a full eigen-decomposition is valid as long as the leading eigenvalues (in terms of absolute value) are in fact the ones driving a superconducting instability.
    In our FRG scheme, the superconducting instability is generated by a divergence stemming from the particle-particle diagrams (the $P$ channel), and stopped only when relatively close to the divergence of $\Gamma^{(4)}$. Therefore, the above condition is satisfied and the eigenvalues corresponding to the superconducting instability become the overall leading eigenvalues of the vertex.%
}%
corresponding to gap functions in the full Brillouin zone and on the Fermi surface, respectively. We note that the TUFRG method is in principle capable of dealing with $\bvec q\not= 0$ instabilities in the particle-particle channel, i.e., pair-density waves.

As next step in our analysis, we decompose the gap function $\hat\Delta(\bvec k)$ into its singlet ($\psi(\bvec k)$) and triplet ($\bvec d(\bvec k)$) components,
\begin{equation}
    \hat\Delta(\bvec k) = \big[ \psi(\bvec k)\mathds 1 + \bvec d(\bvec k)\cdot\hat{\bvec \sigma} \big] \, (i\hat\sigma_y)\ .
\end{equation}
Since inversion symmetry is explicitly broken by the Rashba SOC, a single solution $\hat\Delta(\bvec k)$ may have non-vanishing singlet and triplet components at the same time, i.e., display singlet-triplet mixing. We quantify the degree of singlet-triplet mixing by calculating the absolute weight of the singlet component $\int\!\frac{\dd\bvec k}{V_\mathrm{BZ}}\,\|\psi(\bvec k)\|^2$, which must lay between zero and one. Note that we use the \emph{left} singular vectors for this calculation, as the weights would need to be renormalized when projecting to the Fermi surface.

To obtain further information on the effective pairing interaction, i.e., whether the state is driven by a singlet, triplet or mixed instability, one must explicitly deconstruct the effective pairing interaction $\hat P[\Gamma^{(4)}]_{\bvec{0}fg}^{\sigma_1 \sigma_2 \sigma_3 \sigma_4}$ into odd, even and mixed transformation behavior\,\cite{PhysRevB.77.104520}. In the case of intraband pairing this construction has been extensively discussed in Ref.\,\onlinecite{PhysRevB.77.104520}, however, we have both inter- and intraband pairing and therefore need to resort to the general case of calculating the total singlet (and triplet) weight.

\subsubsection{Particle-hole instabilities}
For instabilities stemming from the crossed or direct particle-hole channels ($C/D$),
we instead calculate the particle-hole susceptibility using the $D$-channel projection of $\Gamma^{(4)}$:
\begin{equation}
    \includegraphics[valign=c]{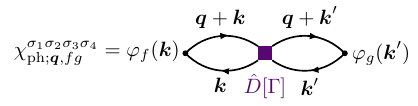}.
    \label{eqn:susc_ph}
\end{equation}
Again, we evaluate the particle-hole loops at the critical scale $\Lambda_\mathrm c$ and diagonalize to find the dominant subspace $\{\bvec q\}$ and the corresponding eigenvectors. The eigenvectors serve as an estimate of the particle-hole gap structure, highlighting the use of this analysis for the different channels. We point out that performing the calculations in spin space -- orbital-spin space for general models -- instead of band space is advantageous due to the lack of gauge invariance of these four-point functions.

Similar to the treatment of particle-particle instabilities, we transform the eigenvector $m_{\sigma\sigma',f}^{\bvec q}$ (with $\bvec q$ the momentum transfer and $f$ the formfactor index) into its charge/spin representation using Pauli matrices,
\begin{equation}
    \hat m_{f}^{\bvec q} = m_{\mathrm{CDW}; f}^{\bvec q} \mathds 1 +
    \bvec m_{\mathrm{SDW}; f}^{\bvec q} \cdot \hat {\bvec \sigma}\,.
\end{equation}
Focusing on the mean-field decoupling of the above magnetic ($\bvec m_{\mathrm{SDW};f}^{\bvec q}$) and charge ($m_{\mathrm{CDW};f}^{\bvec q}$) instabilities we explicitly keep the dependence on a general transfer momentum $\bvec q$. Thereby we can distinguish between ferromagnetic and antiferromagnetic instabilities including incommensurate ordering vectors. The general mean-field Hamiltonian in the particle-hole channel then reads
\begin{equation}
    H_\mathrm{DW}^{\bvec q} = \sum_{\bvec k, f} \big[\varphi_f(\bvec k)\, (\vec c_{\bvec q+\bvec k})^{\dagger} \cdot \hat m_f^{\bvec q} \cdot \vec c_{\bvec k} + \mathrm{H.c.} \big] \,,
    \label{eqn:H_ph}
\end{equation}
where $\vec c_{\bvec k} = (c_{\bvec k,\uparrow}, c_{\bvec k,\downarrow})^\mathrm{T}$.

The particle-hole instabilities found in this work have dominant weight in the trivial formfactor $\varphi_f(\bvec k)\equiv 1$. In this case, time reversal symmetry of the model demands that the scalar (charge) and vector (spin) components must not mix (see \cref{app:trs-derivation}), i.e., either $m_\mathrm{CDW}^{\bvec q}$ or $\bvec m_\mathrm{SDW}^{\bvec q}$ is zero. The non-$SU(2)$ nature of the system may thus only facilitate a possible mixture of the different vectorial spin components. We note that the lack of charge-spin mixing in \cref{eqn:H_ph} differs from the particle-particle case, where the presence of a nontrivial formfactor allows for singlet-triplet mixing.

\section{Results}
\label{sec:results}

\subsection{Phase diagram}

We perform a parameter scan in the three-dimensional parameter-space spanned by next-nearest neighbor hopping $t'\in\{0.0, -0.15, -0.3\}$, Rashba-SOC strength $0 \leq \alpha \leq  0.7$, and filing factor $0.2 \leq \nu \leq 0.8$. We use the nearest-neighbor hopping as energy unit ($t\equiv1$) and set the interaction strength to $U=3$. The wave vectors $\bvec q$ are discretized on a $32\times32$ mesh in the first Brillouin zone, with an additional refinement of $21\times21$ points\,\cite{behak}. The formfactor expansion is truncated after $21$ formfactors, corresponding to fifth nearest neighbors.

\begin{figure}
    \centering
    \includegraphics[width=\columnwidth]{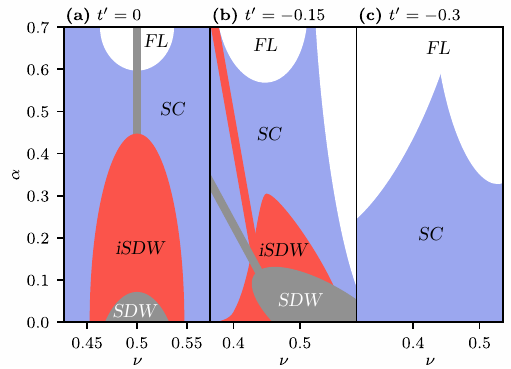}
    \caption{Schematic phase diagram of the square lattice Rashba-Hubbard model as a function of filling factor $\nu$ and Rashba-SOC $\alpha$ for three different nearest-neighbor hopping strengths: $t'=0$~(a), $t'=-0.15$~(b), and $t'=-0.3$~(c). The results were obtained for the weak-to-intermediate coupling regime $U=3$. Superconducting (SC) regions are colored in blue, spin density wave (SDW) instabilities in grey, incommensurate SDWs (iSDW) in red and Fermi liquid (FL) behavior in white.}
    \label{fig:schematic}
\end{figure}

\Cref{fig:schematic} displays a schematic of the resulting phase diagram, where we distinguish between superconducting (SC) and commensurate/incommensurate density wave instabilities (SDW/iSDW) as well as Fermi liquid (FL) like behavior. We observe an intricate interplay between nesting and van Hove singularities for the density wave instabilities which we will elaborate in \cref{sec::ph}. For the superconducting order we observe an increase of singlet-triplet mixing with increasing Rashba coupling $\alpha$, which is discussed thoroughly in \cref{sec::pp}. We do not observe any charge density waves for the considered parameters.

\subsection{Particle-hole instabilities}
\label{sec::ph}

\begin{figure}
    \centering
    \includegraphics[width=\columnwidth]{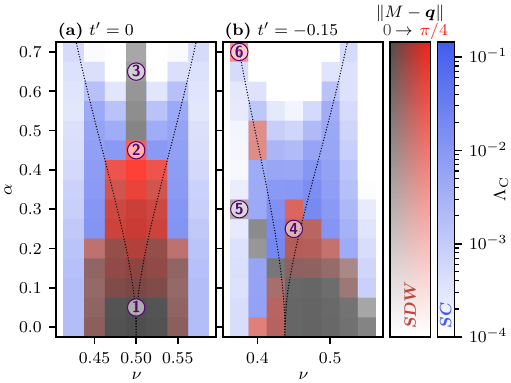}
    \caption{Phase diagram of the Rashba-Hubbard model as a function of filling $\nu$ and SOC strength $\alpha$ at $t'=0$ (a) and $t'=-0.15$ (b). The critical scale $\Lambda_\mathrm c$ roughly corresponds to a transition temperature and is encoded as transparency of the data points. Superconducting phases are encoded in blue, density wave instabilities are classified according to the distance of the leading transfer momentum $\bvec q$ to the $M=(\pi,\pi)$ point of the Brillouin zone: commensurate instabilities ($\|M-\bvec q\|=0$) obtain a gray color and incommensurate ones ($\|M-\bvec q\|>0$) are increasingly red with greater distance to the $M$-point. The position of the van Hove singularities is marked as thin dotted line. Additionally, we annotate the specific points \circled{1}--\circled{6} for which the Fermi surface is visualized in \cref{fig:FS_nesting}.}
    \label{fig:dwinst}
\end{figure}

\Cref{fig:dwinst} presents a more detailed version of the phase diagram (cf.\ \cref{fig:schematic}) focused on (i)SDW order. It is therefore restricted to the cases $t'=0$~(a) and $t'=-0.15$~(b). We here not only encode the type of instability, but also the critical scale $\Lambda_\mathrm c$, which roughly corresponds to the critical temperature associated with the transition, as transparency. Moreover, we continuously color the degree of incommensurability from gray (commensurable, $\bvec q=M\equiv (\pi,\pi)$) to red (incommensurable, $\|\bvec q-M\| = \pi/4$).

\begin{figure}
    \centering
    \includegraphics[width=\columnwidth]{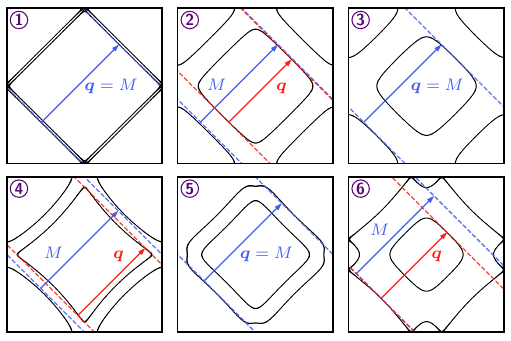}
    \caption{Nesting analysis of SDW instabilities for $t'=0$ (upper row, panels~\circled{1}--\circled{3}) and $t'=-0.15$ (lower row, panels~\circled{4}--\circled{6}). The panels correspond to the points in $\nu,\alpha,t'$ space marked in \cref{fig:dwinst}. The upper row shows the evolution of nesting for increasing $\alpha$ in the case $t'=0$ to explain the return of the commensurate SDW. Blue arrows correspond to $M$-nesting and orange arrows to a $\bvec{q}\neq M$-nesting. In the lower row, we visualize the Fermi-surface and nesting vectors for three selected points in the $t'=-0.15$ phase diagram.}
    \label{fig:FS_nesting}
\end{figure}

Along the vertical line at half filling $\nu=0.5$ in \cref{fig:dwinst}~(a), we observe that upon increasing $\alpha$, the system first is susceptible to commensurate SDW, thereafter to iSDW, and finally again to commensurate SDW order. This effect is explained by the competition between the position of the van Hove singularities and nesting vectors\,\cite{kawano-22arXiv2208.09902}: At low $\alpha$, marked with \circled{1} in \cref{fig:dwinst}, the instability is dominated by perfect nesting of the Fermi surface with respect to $\bvec q=M$, as shown in the upper left panel of \cref{fig:FS_nesting} (panel~\circled{1}). When increasing $\alpha$, the Fermi surface sheets split and the van Hove singularity no longer resides at $X$. This increasingly breaks the $\bvec q=M$ nesting, eventually becoming sub-leading to the new nesting vector between two van-Hove singularities, $\bvec q=M-(\epsilon,\epsilon)^\mathrm{T}$, see \cref{fig:FS_nesting}, panel~\circled{2}. Further increasing $\alpha$, we arrive at a regime where the van Hove singularity is far from the Fermi level, suppressing its influence. Here the ordering vector is determined again by the nesting, now in between the Fermi surfaces as shown in panel~\circled{3}. As the Fermi surfaces are split symmetrically around the $\alpha=0$ diamond, the preferred ordering is again $\bvec q=M$. Because this phase is heavily driven by nesting, small deviations in filling $\nu$ are sufficient to suppress it (see \cref{app:linecuts}).

For $t' = -0.15$ (cf.\ \cref{fig:dwinst}~(b)) we obtain a different picture: Here, when increasing $\alpha$ we have a transition from $\bvec q=M$ to incommensurate ordering vectors. This is explained by the observation that $\bvec{q}$ is the momentum vector between two van Hove points on the inner sheet, while $M$ connects only the center of the arcs (see \cref{fig:FS_nesting}, panel~\circled{4}). The higher density of states at the singularities prevails. The $t'=-0.15$ SDW phase diagram displays two further notable features: First, the commensurate SDW with $\bvec q=M$ extends to lower fillings, eventually becoming a thin line. On this line, the Fermi surface deformation induced by $\alpha$ cancels the one due to $t'$ such that (almost) perfect nesting is recovered on the outer Fermi surface sheets, leading to $\bvec q=M$ SDW order (see panel~\circled{5} of \cref{fig:FS_nesting}). Second, we observe iSDW order for points close enough to the left van Hove singularity. Here, the instability is driven by the divergent density of states with an ordering vector connecting the outer and the inner Fermi surface sheets (see panel~\circled{6} of \cref{fig:FS_nesting}). Due to the finite resolution in filling, our points do not perfectly align with van Hove filling for each $\alpha$ and we see the iSDW order on the left van Hove arm at $t'=-0.15$ only for certain $\alpha$ in the FRG data (cf.\ \cref{fig:dwinst}). Given the above explanation, we expect the feature to prevail for all $\alpha$, indicated accordingly in \cref{fig:schematic}.

\begin{figure}
    \centering
    \includegraphics[width=\columnwidth]{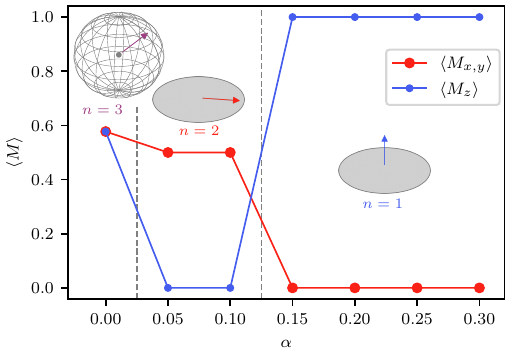}
    \caption{Evolution of possible magnetization vectors in the $\bvec q=M$ phase at $t=-0.15$ (along the line towards point~\circled{5} in \cref{fig:dwinst}~(b)). We show the magnetization direction as an evolution of increasing Rashba coupling $\alpha$. From the initial threefold degeneracy ($n=3$) at $\alpha=0$ we observe an easy-plane SDW (in the $xy$ plane, $n=2$) which turns into an easy axis SDW (along the $z$ axis, $n=1$) for higher $\alpha$.}
    \label{fig:magnetization}
\end{figure}

The distance to commensurate $\bvec q=M$ order serves as primary classification for iSDW phases in \cref{fig:dwinst}. As further analysis, we determine the degeneracy of the maximal eigenvalue of the susceptibility. We find one-, two- and threefold degenerate points, where threefold degeneracy is exclusive to vanishing $\alpha$. The two-fold degenerate points lead to an easy-plane ordering where the magnetization rotates in the $xy$-plane. This property is obtained by transforming the order parameter into real space via $M_i(\bvec r) = |m^{\bvec{q}}_{\mathrm{SDW};i}| \cos\big( \bvec{r}\bvec{q} + \arg(m^{\bvec{q}}_{\mathrm{SDW};i}) \big)$. Since the eigenvalue is strongly peaked at $\bvec q$ the contributions from other ordering vectors can be neglected. An illustration of how the strength of each magnetization component evolves is given in \cref{fig:magnetization}. Here, we vary $\alpha$ for a path entirely within the $\bvec q=M$ ordered phase at $t'=-0.15$ along the line towards point~\circled{5}. We observe a transition from a rotationally symmetric antiferromagnet (AFM) at $\alpha=0$ to an easy-plane ($xy$) AFM at weak $\alpha$ to an easy-axis ($z$) AFM at strong $\alpha$.

\subsection{Particle-particle instabilities}
\label{sec::pp}

\begin{figure}
    \centering
    \includegraphics[width=\columnwidth]{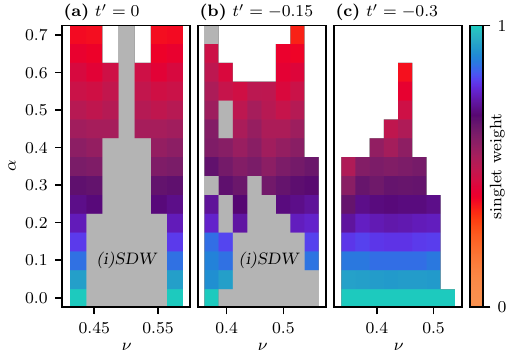}
    \caption{Relative weight of the singlet contribution to the superconducting order parameter for $t'=0$~(a), $t'=-0.15$~(b), and $t'=-0.3$~(c). We observe a purely singlet order parameter at $\alpha=0$, with an increase in mixing as we increase the Rashba coupling for all values of $t'$ under consideration. Note that the (i)SDW phases are grayed out for visual clarity. We do not observe a strong dependence of the singlet-triplet mixing on the filling $\nu$ for any value of $\alpha$.}
    \label{fig:scinst}
\end{figure}

We now turn our attention to the analysis of the superconducting instabilities shown in \cref{fig:schematic}. As discussed in \cref{sec:modelmethod} we transform the superconducting gap function from spin space to its singlet and triplet components. We show the resulting relative singlet weight in \cref{fig:scinst}. Additionally, we determine the irreducible representation of the order parameter. Note that by construction both the singlet and triplet components must transform in the same irreducible representation. We therefore can resort to barely analyzing the singlet component, finding B1 ($d_{x^2-y^2}$-wave) for all superconducting instabilities. In the triplet channel, the spin itself transforms as a pseudovector, meaning that instead of an  B1 irreducible representation in momentum space, we expect an E irreducible representation.

\Cref{fig:scinst} reveals that the relative weight of the singlet component, which serves as an indicator for the strength of singlet-triplet mixing, almost linearly depends on the Rashba-SOC strength, with slight saturation effects at high $\alpha$. We emphasize that at values as high as $\alpha=0.7$ we obtain more than 50 percent triplet contribution in the superconducting ground state, which can be of the $p+ip$ type and, in principle, give rise to helical topological superconductivity.
At the $SU(2)$ symmetric points ($\alpha=0$) we observe the expected Hubbard-model behavior of singlet $d_{x^2-y^2}$-wave superconductivity.
The B1-representation is in our case equivalent to an 
$d_{x^2-y^2}$ superconductor in the singlet, or an admixture between $p$ and $f$-wave in the triplet component. 

\section{Conclusion}
\label{sec:conclusion}

We firmly establish the truncated unity functional renormalization group as a method to study strongly correlated few-orbital systems with spin-orbit coupling. We add a Rashba-type spin-orbit interaction to the paradigmatic square lattice Hubbard model and make two main observations.

First, our calculations reveal that the the FRG phase diagram is stable against small values of Rashba-SOC $\alpha$. For weakly spin-orbit coupled systems, the correlated phases only experience slight changes: The AFM order gives way to incommensurate SDWs for certain parameter sets, and superconducting instabilities acquire weak singlet-triplet mixing.

Second, we uncover a richer phenomenology for systems with larger values of $\alpha$. There, we find a delicate interplay of $t'$ and $\alpha$, which leads to accidental nesting, resulting in commensurate AFM phases. Moreover, we observe a competition between nesting- and van-Hove-driven (i)SDW order. The superconducting instabilities develop singlet-triplet mixing roughly proportional proportional to $\alpha$, showing no strong dependence on filling $\nu$.

\begin{acknowledgements}
The German Research Foundation (DFG) is acknowledged for support through RTG 1995, within the Priority Program SPP 2244 \enquote{2DMP} and under Germany's Excellence Strategy-Cluster of Excellence Matter and Light for Quantum Computing (ML4Q) EXC20004/1-390534769.
R.T. acknowledges support from the DFG through QUAST FOR 5249-449872909 (Project P3), through Project-ID 258499086-SFB 1170, and from the W\"urzburg-Dresden Cluster of Excellence on Complexity and Topology in Quantum Matter – ct.qmat Project-ID 390858490-EXC 2147. 
S.R. acknowledges support from the Australian Research Council (FT180100211 and DP200101118).
The authors gratefully acknowledge the scientific support and HPC resources provided by the Erlangen National High Performance Computing Center (NHR@FAU) of the Friedrich-Alexander-Universität Erlangen-Nürnberg (FAU) during the NHR@FAU early access phase. NHR funding is provided by federal and Bavarian state authorities. NHR@FAU hardware is partially funded by the DFG – 440719683.
\end{acknowledgements}

\appendix

\section{Validation of results}
\label{sec:validation}

\begin{figure}
    \centering
    \begin{tikzpicture}
    \node[anchor=center] at (0,0) {\includegraphics[width=\columnwidth]{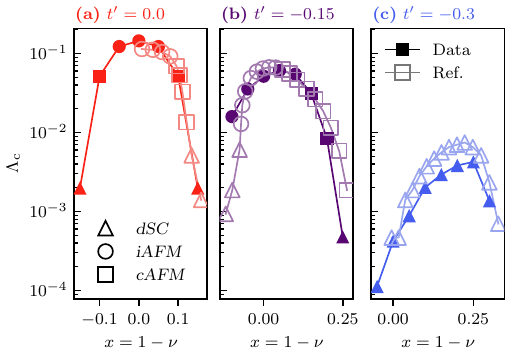}};
    \node[anchor=north west] at (2.89,2.01) {\footnotesize{}\phantom{Ref.}~\onlinecite{PhysRevB.89.035126}};
    \end{tikzpicture}
    \caption{Comparison to Ref.\,\onlinecite{PhysRevB.89.035126} for $\alpha=0$. Note that we have globally rescaled the critical scales of our data. This global offset is easily explained by the differences in implementations, the reference used a sharp cutoff as well as constant self-energies, wich are absent here. It is nevertheless apparent that the results have the same features, with slightly shifted transition fillings between phases. cAFM (iAFM) corresponds to our SDW (iSDW) phase and dSC stands for $d$-wave superconductivity.}
    \label{fig:comparison_metzner}
\end{figure}

The code used to generate these results is one of three in the recently published equivalence class of FRG codes\,\cite{behak}. We therefore refer the interested reader to that publication for validation, the binary equivalence shown there is a stronger indicator than what could be provided here. Nevertheless we want to show agreement with previous calculations performed on the square lattice Hubbard model in Ref.\,\onlinecite{PhysRevB.89.035126}. The calculations performed there align with the $\alpha=0$ results of our phase diagrams in \cref{fig:dwinst}, as we demonstrate in \cref{fig:comparison_metzner}. Our data show good agreement with their results in critical scale, the transition fillings are shifted slightly. Note however that a global rescaling of the (arbitrary) critical scale was performed to obtain this match. These minor differences are due to chosen regulator schemes and self-energy treatment. They omitted the particle-hole symmetric half of the $t'=0$ case, we choose to show that it is indeed symmetric.

\section{Mixing of SDW and CDW}
\label{app:trs-derivation}

In this appendix, we show that for particle-hole instabilities with on-site formfactor ($\varphi_f(\bvec k)\equiv 1$), charge- and spin-sectors of the density waves cannot mix. The on-site formfactor implies that the order parameters in \cref{eqn:H_ph} are independent of $\bvec{k}$. Since the Hamiltonian has to be Hermitian even in the symmetry broken phase, we obtain
\begin{equation}
\begin{aligned}
    H_\mathrm{DW}^\dagger
    &{}= \sum_{\bvec q,\bvec k,\nu} \left( \vec{c}_{\bvec k+\bvec q}^{\,\dagger} \cdot \hat \sigma^\nu \cdot \vec{c}_{\bvec k}^{\vphantom{\dagger}} \, m^{\bvec q}_\nu \right)^\dagger \\
    &{}= \sum_{\bvec q,\bvec k,\nu} \vec{c}_{\bvec k}^{\,\dagger} \cdot \hat\sigma^\nu \cdot \vec c_{\bvec k+\bvec q}^{\vphantom{\dagger}}\, (m_\nu^{\bvec q})^* \\
    &{}= \sum_{\bvec q,\bvec k,\nu} \vec{c}_{\bvec k+\bvec q}^{\,\dagger} \cdot \hat\sigma^\nu \cdot \vec c_{\bvec k}^{\vphantom{\dagger}} \, (m_\nu^{-\bvec q})^* \\
    &{}\stackrel{!}{=} H_\mathrm{DW}
\end{aligned}
\end{equation}
i.e., $(m^{-\bvec q}_\nu)^{*} = m^{\bvec q}_\nu$ (for $\nu=0,x,y,z$). Applying time reversal $\mathcal T$ to the general density-wave Hamiltonian yields
\begin{equation}
\begin{aligned}
    \mathcal T H_{\text{DW}} \mathcal T^{-1}
    &{}= \sum_{\bvec q,\bvec k,\nu} \big(\hat\sigma^y\cdot\vec{c}_{-\bvec q-\bvec k}\big)^\dagger \cdot \big(\hat{\sigma}^\nu \, m^{\bvec q}_\nu\big)^* \cdot \big(\hat\sigma^y \cdot \vec{c}_{-\bvec k}\big)\\
    &{}= \sum_{\bvec q,\bvec k,\nu} \vec{c}_{-\bvec q-\bvec k}^{\,\dagger} \cdot \big( \hat\sigma^y (\hat\sigma^\nu)^* \hat\sigma^y \big) \cdot \vec{c}_{-\bvec k}^{\vphantom{\dagger}} \, (m_\nu^{\bvec q})^* \\
    &{}= \sum_{\bvec q,\bvec k,\nu} \vec{c}_{\bvec k+\bvec q}^{\,\dagger} \cdot \big( \hat\sigma^y (\hat\sigma^\nu)^* \hat\sigma^y \big) \cdot \vec{c}_{\bvec k}^{\vphantom{\dagger}} \, (m_\nu^{-\bvec q})^* \\
    &{}= \sum_{\bvec q,\bvec k,\nu} \vec{c}_{\bvec k+\bvec q}^{\,\dagger} \cdot \hat\sigma^\nu \cdot \vec{c}_{\bvec k}^{\vphantom{\dagger}} \, m_\nu^{\bvec q} \, \eta_\nu \,,
\end{aligned}
\end{equation}
with $\eta_0 = +1$ (i.e., the charge component; $\hat\sigma^y(\hat\sigma^0)^*\hat\sigma^y=\hat\sigma^0$) and $\eta_{x,y,z}=-1$ (i.e., the spin components; note that $\hat\sigma^y(\hat\sigma^{i})^*\hat\sigma^y=-\hat\sigma^i$ for $i=x,y,z$). We split the Hamiltonian into its charge- and spin sectors,
\begin{equation}
\begin{aligned}
    H_\mathrm{DW} &{}= H_\mathrm{CDW} + H_\mathrm{SDW} \\
    &{}= \begin{multlined}[t]
        \sum_{\bvec q,\bvec k} \vec c_{\bvec k+\bvec q}^{\,\dagger} \cdot \vec c_{\bvec k}^{\vphantom{\dagger}} \, m_\mathrm{CDW}^{\bvec q}
        + {}\\
        \sum_{\bvec q,\bvec k}
        \vec c_{\bvec k+\bvec q}^{\,\dagger} \cdot (\hat{\bvec \sigma}\cdot \bvec m_\mathrm{SDW}^{\bvec q}) \cdot \vec c_{\bvec k}^{\vphantom{\dagger}} \,,
    \end{multlined}
\end{aligned}
\end{equation}
which results in the following transformation behavior under $\mathcal T$:
\begin{align}
    \mathcal{T} H_\mathrm{CDW} \mathcal{T}^{-1} &{}= +H_\mathrm{CDW} \,, \\
    \mathcal{T} H_\mathrm{SDW} \mathcal{T}^{-1} &{}= -H_\mathrm{SDW} \,.
\end{align}
Thus, the subspaces of $H_\mathrm{SDW}$ and $H_\mathrm{CDW}$ are orthogonal. If the initial Hamiltonian (without symmetry breaking) is invariant under $\mathcal T$ (as it is for the Rashba-Hubbard model), the SDW and CDW phases belong to different irreducible representations of $\mathcal T$ and therefore, a mixing is forbidden.

\section{Linecuts through phase diagrams}
\label{app:linecuts}

\begin{figure*}
    \centering
    \includegraphics[width=\textwidth]{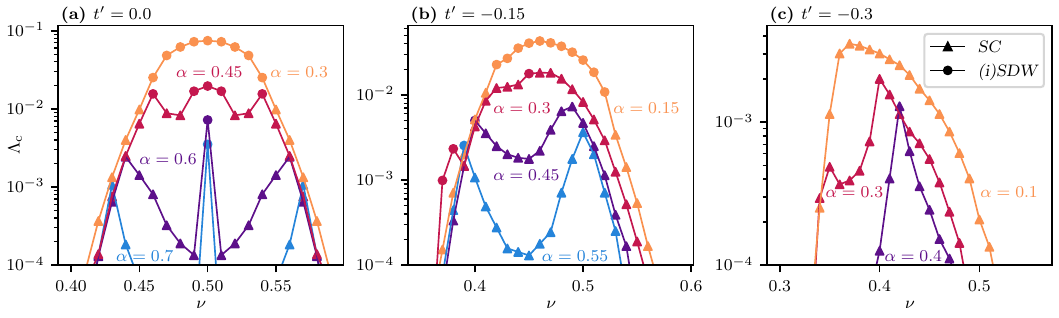}
    \caption{We show linecuts through the plots of \cref{fig:dwinst,fig:scinst} with an increased resolution in the filling $\nu$. This serves to both confirm the main features of the above calculation as well as establish the width of the nesting lines.}
    \label{fig:linecuts}
\end{figure*}

To obtain an improved understanding of the features at points \circled{3} and \circled{5} of \cref{fig:dwinst}, we calculate slices through the phase diagram intersecting the features. The resulting lines can be seen in \cref{fig:linecuts}. We want to place special emphasis on the lines for $\alpha = 0.6$ (purple) and $\alpha = 0.7$ (blue); see panel~(a) ($t'=0$) as well as the line for $\alpha = 0.3$ (red) in panel~(b) ($t'=-0.15$). Here we can distinguish between \emph{true} nesting of the former and the accidental nesting of the latter. While the magnetism in the high-$\alpha$ regime of $t'=0$ is shown to be very narrow (in our spacing exactly one parameter point in width), the nesting for the aforementioned $t'=-0.15$ case is instead created in a parameter region where the effects on the Fermi surface cancel. This is indicated by the width of the magnetic instability in the red line of panel (b).

\bibliographystyle{apsrev4-2}
\bibliography{references}

\end{document}